# Defect Prediction of Railway Wheel Flats based on Hilbert Transform and Wavelet Packet Decomposition

Euiyoul Kim [a,1], Nithya Jayaprakasam [a,2], Yong Cui [a,b,3,*], Ullrich Martin [a,4]

[a] Institut für Eisenbahn- und Verkehrswesen, Universität Stuttgart
Pfaffenwaldring 7, 70569 Stuttgart, Germany
[b] Hefei University
373 Huangshan Rd, Shushan Qu, Hefei Shi, Anhui Sheng, China, 230031

[1] E-mail: euiyoul.kim@cdfeb.de, Phone: +49 (0) 711 685 66362
[2] E-mail: Nithy.civil@gmail.com, Phone: +49 (0) 1521 6733428
[3] E-mail: yong.cui@ievvwi.uni-stuttgart.de, Phone: +49 (0) 711 685 66362
[4] E-mail: ullrich.martin@ievvwi.uni-stuttgart.de, Phone: +49 (0) 711 685 66367
[*] Corresponding author

## Abstract

For efficient railway operation and maintenance, the demand for onboard monitoring systems is increasing with technological advances in high-speed trains. Wheel flats, one of the common defects, can be monitored in real-time through accelerometers mounted on each axle box so that the criteria of relevant standards are not exceeded. This study aims to identify the location and height of a single wheel flat based on non-stationary axle box acceleration (ABA) signals, which are generated through a train dynamics model with flexible wheelsets. The proposed feature extraction method is applied to extract the root mean square distribution of decomposed ABA signals on a balanced binary tree as orthogonal energy features using the Hilbert transform and wavelet packet decomposition. The neural network-based defect prediction model is created to define the relationship between input features and output labels. For insufficient input features, data augmentation is performed by the linear interpolation of existing features. The performance of defect prediction is evaluated in terms of the accuracy of detection and localization and improved by augmented input features and highly decomposed ABA signals. The results show that the trained neural network can predict the height and location of a single wheel flat from orthogonal energy features with high accuracy.

## Keywords

## Nomenclature

| | |
|---|---|
| $h\ [m]$: | Height of a wheel flat |
| $l\ [m]$: | Length of a wheel flat |
| $r_w\ [m]$: | Radius of a wheel, a constant value of 0.5 m |
| $r_{min}\ [m]$: | Minimum radius of a wheel |
| $\theta\ [rad]$: | Angle of a wheel flat |
| $H(x(t))$: | Hilbert transform of real-valued signal |
| $t$: | Time point |
| $\tau$: | Dummy variable |
| $x_p(t)\ [m/s^2]$: | Real-valued signal |
| $x_p^A(t)\ [m/s^2]$: | Complex-valued signal |
| $A_p(t)\ [m/s^2]$: | Analytic amplitude |
| $p$: | Measurement location within one bogie |
| $\emptyset(t)\ [rad]$: | Analytic phase, a constant value of $\pi/2$ |
| $W_{j,k}^n$: | Wavelet packet function |
| $d_{j,k}^n$: | Wavelet packet coefficient |
| $j$: | Scale parameter |
| $m$: | Modulation parameter |
| $k$: | Translation parameter |
| $g(k)$: | Low-pass filter |
| $h(k)$: | High-pass filter |
| $N$: | Length of the Daubechies wavelets |

# 1   Introduction

In the railway industry, structural health monitoring systems are important to efficiently support the decision-making process of maintenance scheduling algorithms regarding safety and cost. Otherwise, trivial defects can cause the sudden suspension of railway operations or even terrible accidents, such as the derailment of a high-speed train, but these potential problems can be efficiently prevented and managed through advanced defect detection methods and well-organized maintenance scheduling.

Among various defects, defects on wheels and axles are the main causes of railway accidents [1, 2]. EN 15313 classified wheel defects into six groups as follows: flange tip, plate failures, scattered rim defects, notching, thermal cracks, and defects on the tread surface [3]. Statistically, one of the most common defects is a wheel flat, which belongs to the group of defects on the tread surface [4]. Wheel flats occur as localized surface defects due to local deformation and damage when wheels are locked by braking or sliding on the rail [5]. The shape of a newly formed wheel flat has sharp edges and soon turns into rounded edges due to wear and plastic deformation [6]. Wheel flats that exceed the criteria of relevant standards should be properly maintained at a balance between safety and cost. EN 15313 defines the maximum length of a wheel flat between 20 and 80 mm depending on wheel diameter, axle load, and train speed [3]. Existing studies generally consider a skid length of 50 to 60 mm and a flat height 0.9 to 1.4 mm to be the allowable limits [7, 8].

Inspections for the maintenance of a wheel flat are classified into in-workshop and in-service inspections according to measurement conditions, and the in-service inspection is further divided into wayside and onboard inspections [9]. The in-workshop and wayside inspections use non-destructive testing equipment that is fixed or has limited movement at a specific location, and one piece of testing equipment can inspect multiple structural parts one-by-one using ultrasonic, magnetic, vision, and fibre optic testing procedures. The onboard inspection can only monitor structural parts adjacent to sensors in a train and requires a relatively large number of sensors, but the advantage is that it can continuously monitor defects during railway operations before and after in-workshop and wayside inspections. For efficient railway operation and maintenance, recent trends and technological advances in high-speed trains place more demands on the onboard health monitoring systems than in the past [10].

During railway operations, micro-electro-mechanical systems (MEMS) capacitive or piezoelectric accelerometers mounted on each axle box are usually chosen as onboard sensors for the detection of wheel flats [11]. The relation between wheel flats and the resulting axle box acceleration (ABA) signals should be studied first using experimental approaches or numerical approaches. With experimental approaches, there are limitations in considering all the necessary defects and operating conditions due to cost and time. As an alternative, numerical approaches are widely used to study the effects of the various defects and operating conditions on ABA signals. Numerical approaches include three types: numerical simulation, feature extraction, and defect prediction.

As the basis of numerical approaches, numerical simulations can simulate railway vehicle dynamics and generate vibration responses in the form of signals at desired positions. Numerical models have become more realistic because of improved wheel-rail contact models [12-14], multiple flats [8], coupling of multibody dynamics (MBD) and finite element methods (FEMs) [15-17], with range expansion of models from train to infrastructure [18, 19]. In this work, because of the difference in influence between rigid and flexible wheelsets on vibration responses [17], a train dynamics model with flexible wheelsets, including structural resonance and damping, will be built to generate ABA signals due to single wheel flats.

The objective of feature extraction is to generate more representative and manageable data for defect prediction from the original ABA signals. The common feature extraction methods are classified into three domains: time, frequency, and time-frequency [20, 21]. Since a non-stationary ABA signal due to a single wheel flat has motion in both time and frequency domains simultaneously, considering only one domain may be the loss and distortion of defect information. Therefore, existing studies have been used to extract statistical values from time-frequency analysis methods based on Fourier transform, wavelet transform, Wigner-Ville transform, etc., using different types of basis functions [22-24]. Signal decomposition methods, such as empirical mode decomposition (EMD) and wavelet decomposition (WD), have also been used to enhance the signal-to-noise ratio (SNR) or to extract the pattern information from orthogonally decomposed subspaces on a binary tree structure [25, 26]. In this work, the combination of Hilbert transform (HT) and wavelet packet decomposition (WPD) will be used to orthogonally extract decomposed signals from the amplitude modulation part that is considered to be closely related to changes in flat height from ABA signals [26, 27]. The extracted features will be given in the form of the root mean square (RMS) distribution on a balanced binary tree for detect prediction.

The goal of this study is to identify the height and location of a single wheel flat from non-stationary ABA signals through a defect prediction model. Recently, the applications of machine learning and deep learning are rapidly increasing to create more accurate defect prediction models [25, 28]. The relationship between input features and output labels will be modelled by a feedforward neural network (FNN). Finally, the height of a wheel flat in mm, the accuracy of detection in %, the location of bogie, wheelset, and wheel with a wheel flat, and the accuracy of localization in % can be inferred from the collected ABA signals.

The remainder of the article is organized as follows. Section 2 describes the modelling and simulation process of a train dynamics model. Section 3 introduces the proposed feature extraction method in detail. In Section 4, the defect prediction process is discussed for better performance. The conclusions are given in Section 5.

## 2  Modelling and Simulation

### 2.1  Train Dynamics Model

A train dynamics model with flexible wheelsets was created to obtain ABA signals from axle box accelerometers due to wheel flats. SIMULIA Simpack Rail is used for modelling and simulation. Figure 1 shows the schematic lateral view of a train dynamics model. The main structure is a vehicle body mounted on the front and rear bogies, where a bogie consists of frame, wheelsets, axle boxes, and suspensions. The wheelset has a hollow axle and two wheels. A wheel with a radius of 0.5 m was chosen. The lateral wheel distance is 0.75 m. The wheel profile of UIC S1002 is used. The axle boxes are located at both ends of a wheelset and include virtual accelerometers inside to measure ABA signals. The primary suspension connects axle boxes with a frame. The secondary suspension is located at the top centre of a frame between bogie and vehicle body. Both suspensions consist of spring and damper in parallel. Rigid bodies are assembled into a train dynamics model with force and joint elements. The wheel-rail contact model is necessary to describe normal and tangential forces at contact points between wheel and rail, which are calculated with Hertzian contact and FASTSIM models, respectively [12]. The rail is modelled with single profile of UIC 60 with a rail inclination of 1:40. In this work, the track and rail are simply modelled to exclude their effects on the vibration response of a train dynamics model. The track is straight and does not include the effects of super-elevation and irregularity that cause vertical and lateral excitation. Starting from the ideal situation, the vibration caused only by wheel flats can be studied without noises. The approaches and findings can serve as the basis for a more complicated situation that considers the effects of wheel flats on the vibration response.

Based on the train dynamic model, randomly generated wheel flats can be introduced. The model of a wheel flat will be described in Section 2.2. In addition, the flexibility of a wheelset is important because of the effect of structural resonances on the wheel-rail interaction force, and this flexibility affects the vibration response in the axle box [17]. In Section 2.3, the procedures to integrate the finite element model of a wheelset into the train dynamics model are explained. The simulation results are presented in Section 2.4.

**2.2   Wheel Flat Model**

The dynamic behaviour of the wheel-rail contact model highly depends on the curvature of a wheel as a function of the arc-length. Discontinuities along the wheel profile can cause non-stationary vibration responses. Figure 2 shows the geometry of a wheel flat with sharp edges, which is analytically modelled in the function of flat height or skid length [29].

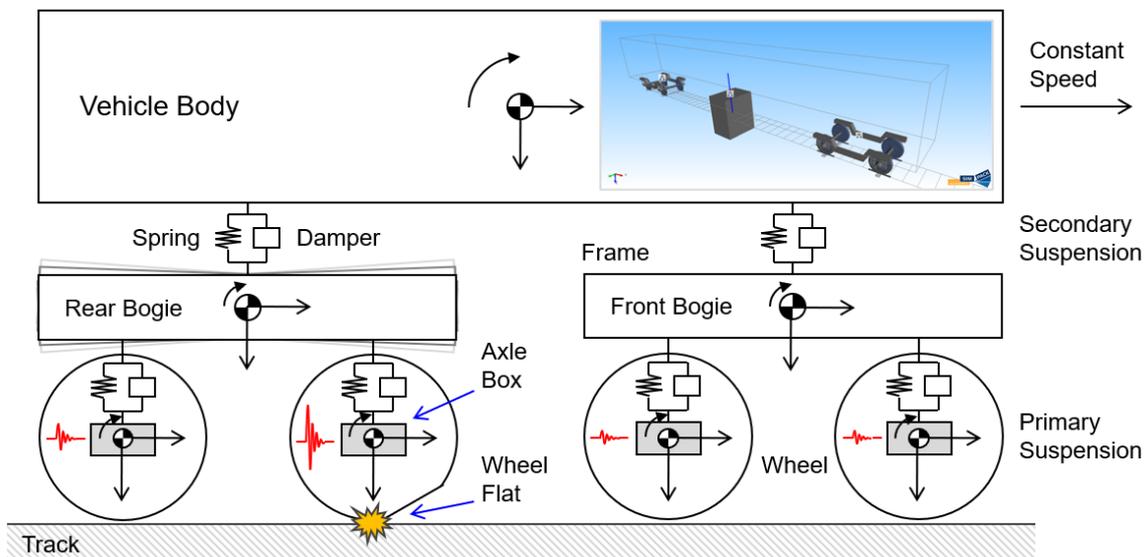

Figure 1 Schematic representation of a train dynamics model with flexible wheelsets and a single wheel flat

The flat height $h$ is defined as the difference between the original radius $r_w$ of a wheel and the minimum radius $r_{min}$ at the centre of a wheel flat with a cosine function, as below:

$$h = r_w - r_{min} = r_w - r_w \cos\theta = r_w(1 - \cos\theta) \qquad \text{Eq. 1}$$

and the skid length $l$ is defined as the distance between the front and back ends of a wheel flat and can be expressed by:

$$l = 2r_w \sin\theta \qquad \text{Eq. 2}$$

If rounded edges are used instead of sharp edges, the magnitude of non-stationary vibration responses will be reduced compared to that of sharp edges because of reduced discontinuities between wheel and rail by wear and plastic deformation [6]. Although the vibration response is likely to be overestimated due to sharp edges, to simplify the geometry of a wheel flat, only sharp edges are considered during the modelling and simulation process.

In general, existing studies consider a skid length of approximately 60 mm and a flat height of 0.9 to 1.4 mm as a critical point according to criteria for the management of wheel flats. A list of evenly divided skid lengths between 0 and 60 mm is commonly used for numerical and experimental applications [7, 8]. In this study, the geometry of a wheel flat is defined as a function of height. The original radius $r_w$ is 0.5 m. A flat height of 1e-0 mm is chosen as a critical point and corresponds to a skid length of 63 mm. The lowest flat height is 1e-4 mm and corresponds to a skid length of 0.2 mm. The range of a flat height is determined in 5 levels between 1e-4 mm and 1e-0 mm and biased on the logarithmic scale towards the lower value. The lower range is emphasized to evaluate the limitations of the proposed energy-based feature extraction and defect prediction in terms of the accuracy of detection and localization. For the efficient modelling for a stochastic approach, the generation process of a wheel flat in SIMULIA Simpack Rail is fully automated with Python. The above range of a flat height is used to calculate the wheel profile between 0 and $2\pi$ in the circumferential direction.

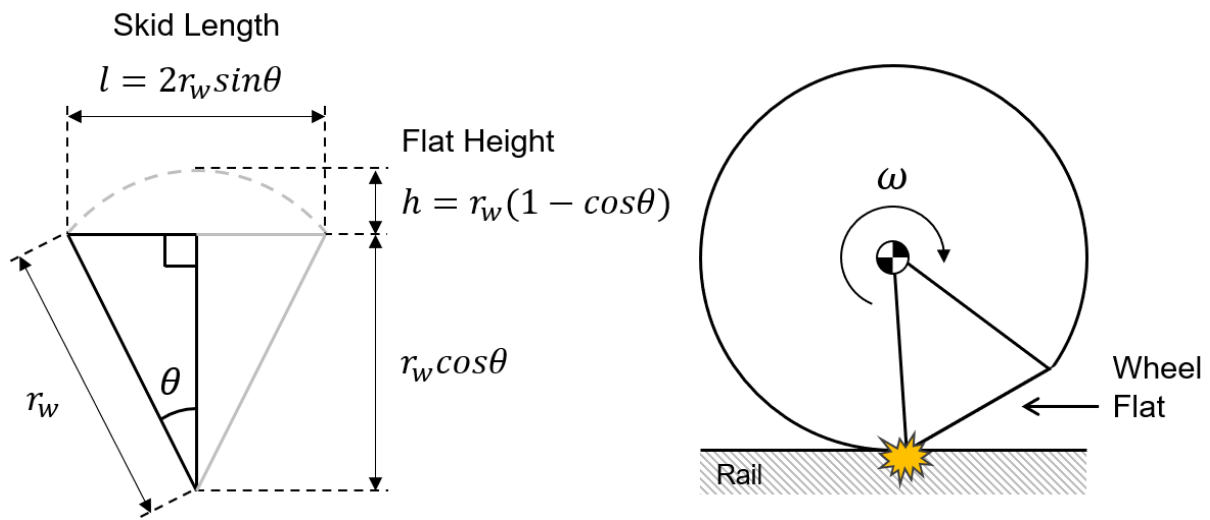

Figure 2 Geometry of a wheel flat with sharp edges

## 2.3 Discretization and Dynamic Reduction of a Flexible Wheelset Model

A train dynamics model is commonly coupled with finite element models to obtain more realistic ABA signals, and the range of coupling is also extended from a wheelset to infrastructure [18, 19]. In this study, the flexibility of a wheelset is considered in the modelling and simulation. Modal properties are used to describe vibration response characteristics of a flexible wheel with natural frequency, mode shape, and damping ratio. The integration process from a flexible wheelset to a train dynamics model consists of four steps as follows: Computer-aided design (CAD) modelling, FE discretization, modal analysis, and dynamic reduction.

Step 1) The geometry of a wheelset is created by using AutoCAD and consists of one hollow axle and two wheels with the wheel profile of UIC S1002.

Step 2) The finite element model of a wheelset is created using only hexahedral elements with the linear shape function and reduced integration (C3D8R). The discretization is repeatedly performed with HyperMesh until numerical errors that cause parasitic structural stiffness, such as element locking, are sufficiently minimized. Among the modal properties, the change in natural frequencies is used as a criterion to evaluate the convergence of a finite element model.

Step 3) The modal properties of a flexible wheelset are estimated under free-free conditions by using the Lancos Eigen solver for the modal analysis in Abaqus/Standard. The sampling rate of a transient simulation in SIMULIA Simpack Rail is 2,000 Hz, and the bandwidth is 1,000 Hz according to Nyquist's rule [30]. The frequency range of the modal analysis is determined to include the mode shapes of interest and the stiffness line of resonances located in the higher frequency range. Figure 3 shows the flexible mode shapes and natural frequency of a flexible wheelset that are considered important in related studies [15-17]. The modes from 1 to 6 are rigid body modes. Thus, the frequency range of the modal analysis in Abaqus/Standard is chosen between 0 and 1,500 Hz with a margin of 50%. The estimated modal properties include more resonance modes up to 1,500 Hz, which can be seen in Figure 3.

Step 4) The degrees of freedom of a finite element model are minimized for efficient simulations without the loss of modal properties. Using all the degrees of freedom for a discretized wheelset model is inefficient in terms of computational cost and time. The Craig-Bampton method [31] is used to transfer the modal properties of interest from a full model to a reduced model using a total of 57 flexible markers. A flexible marker is a connection point between full and reduced models. Figure 4 shows that the reduced model consists of two wheels of 16 flexible markers that are equally spaced at the end of the wheel radius, and an axle of 25 flexible markers. The number and location of flexible markers need to be carefully chosen to minimize spatial aliasing problems depending on the complexity of mode shapes. Otherwise, shock vibrations occur periodically due to a mismatch between the full and reduced models. As a result, the degrees of freedom are reduced from 19,938 DoF in a full model to 171 DoF in a reduced model with a damping ratio of 2% and then are integrated into wheelsets of a train dynamics model. The vibration response characteristics of structural transfer paths from an excitation point to left and right axle boxes are replaced by a reduced model.

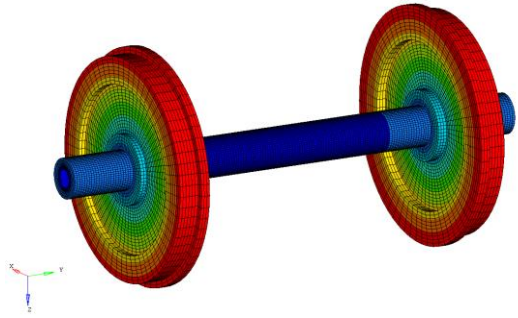

(a) Mode 7 (55.626 Hz)
First torsion mode of the axle

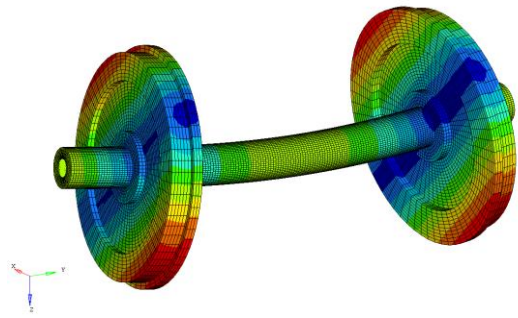

(b) Mode 8 (76.292 Hz)
First bending mode of the axle

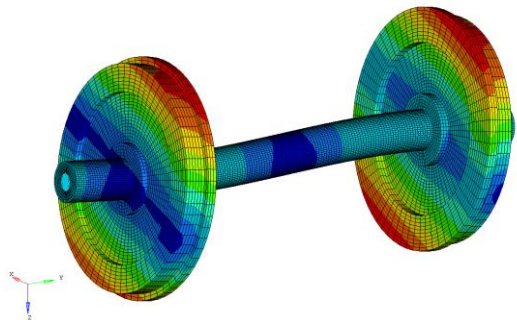

(c) Mode 9 (136.996 Hz)
Second bending mode of the axle

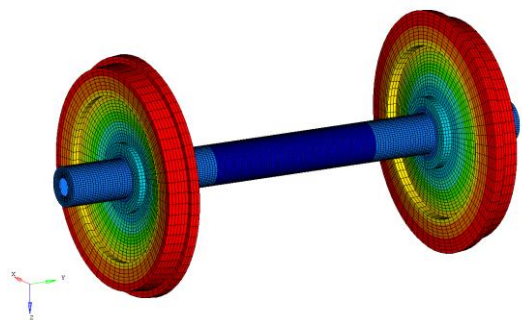

(d) Mode 10 (279.376 Hz)
First umbrella mode of the wheel

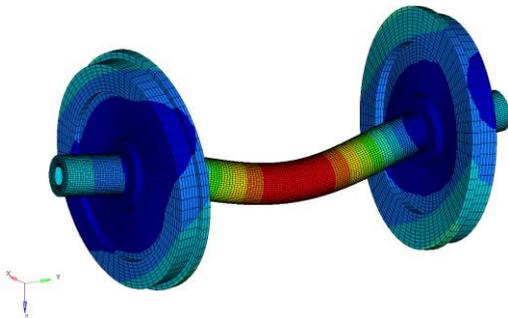

(e) Mode 11 (365.859 Hz)
First bending mode of the wheel and axle

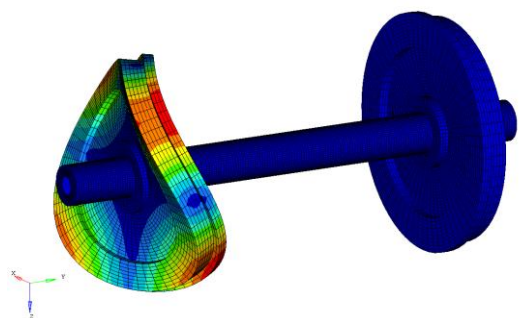

(f) Mode 12 (445.490 Hz)
First bending mode of the wheel

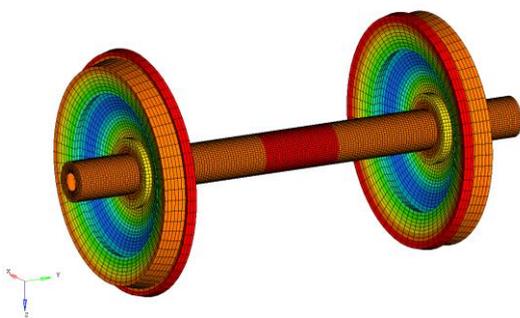

(g) Mode 13 (445.490 Hz)
Second umbrella mode of the wheel

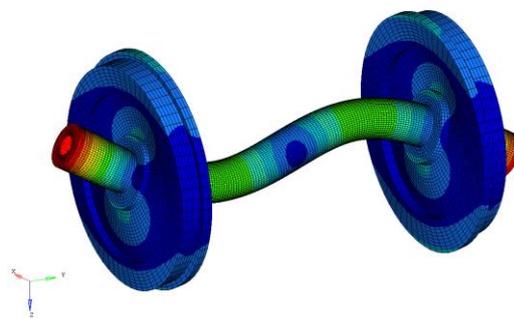

(h) Mode 14 (731.255 Hz)
Second bending of the wheel and axle

Figure 3 Mode shapes of a discretized wheelset model

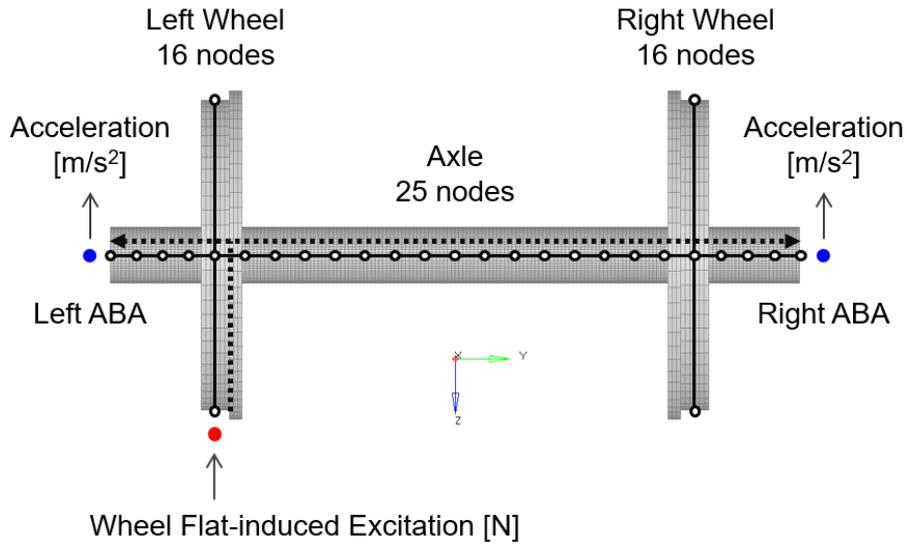

Figure 4 Dynamic reduction and excitation energy flow in a discretized wheelset model

## 2.4 Simulation and Discussion of Results

The train dynamics model is assumed to be running on the straight track at a constant speed of 60 km/h. Wheel flats are randomly generated as a function of flat height between 1e-4 mm and 1e-0 mm and are located on one of the four wheels in the front bogie. The dynamic behaviour is calculated at the sampling rate of 2,000 Hz by the default solver, SODASRT2, in SIMULIA Simpack Rail [32], and then, the ABA signals measured at each axle box are stored in the CSV format.

Figure 5 shows simulated non-stationary ABA signals due to a flat height of 1e-1 mm by the location of ABAs in the front bogie. A single wheel flat is located at the front wheelset and left wheel. Although there is a slight difference in signal shape for each wheel rotation, the difference between signals is statistically consistent in terms of magnitude and phase. The attenuation and delay of vibrational energy depend on structural transfer paths from an excitation point to axle box accelerometers. The difference between the front and rear wheelsets is large enough to intuitively predict the location of a wheel flat. In both wheels of the rear wheelset, the magnitude is smaller than those of the front wheelset, and there is a delay of approximately 120 degrees in the wheel angle. In the front wheelset, the difference between the left and right wheels is relatively small, and there is little delay because vibrational energy is transferred directly through the structure of a wheelset. Figure 6 shows that the vibrational energy transferred from the front bogie to the rear bogie is very small because structure transfer paths are longer and contain more damping elements than are considered only within one bogie. Thus, the vibrational energy caused by a wheel flat is not easily transferred to other bogies due to high structural damping. Therefore, the proposed energy-based methodology for detecting wheel flats is applied only within one bogie. When a flat height increases from 1e-1 mm to a critical value of 1e-0 mm, the severe non-contact between wheel and rail constantly appears to be a jump, and the magnitude of non-stationary ABA signals increases dramatically up to approximately max ±100 g. The range of flat heights considered in the simulation is covered within the general range, ± 200 g, of MEMS capacitive accelerometers for axle boxes [33].

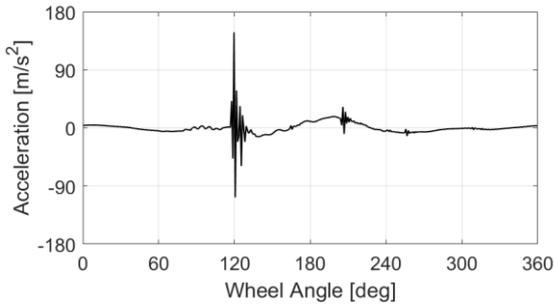
(a) Front wheelset-Left wheel (Defect)

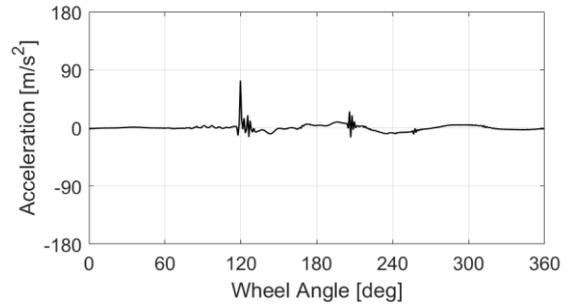
(b) Front wheelset-Right wheel

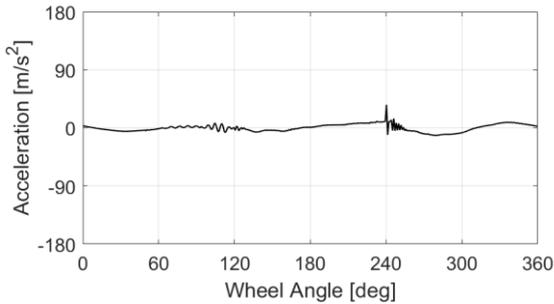
(c) Rear wheelset-Left wheel

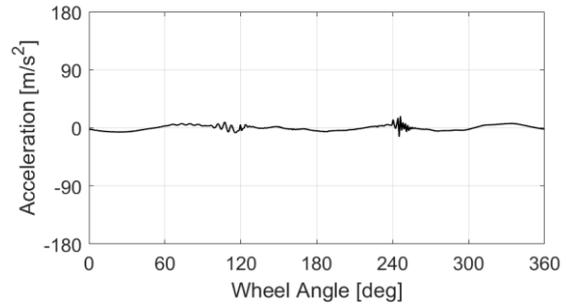
(d) Rear wheelset-Right wheel

Figure 5 Comparison of simulated vertical ABA signals due to a flat height of 1e-1 mm by the location of axle box accelerometers in the front bogie

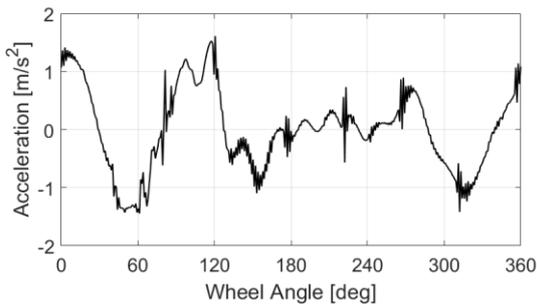
(a) Front wheelset-Left wheel

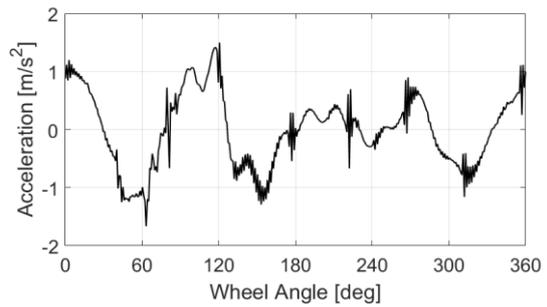
(b) Front wheelset-Right wheel

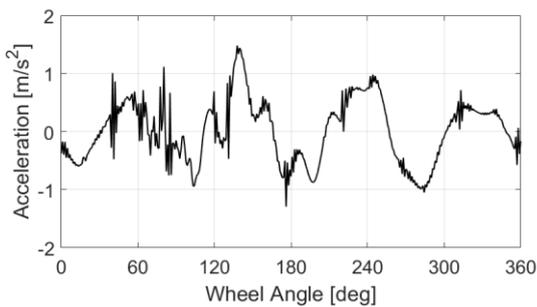
(c) Rear wheelset-Left wheel

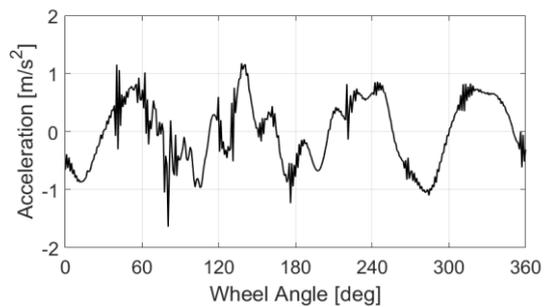
(d) Rear wheelset-Right wheel

Figure 6 Comparison of simulated vertical ABA signals due to a flat height of 1e-1 mm by the location of axle box accelerometers in the rear bogie

# 3 Feature Extraction

## 3.1 Proposed Energy-based Methodology

The proposed methodology aims to extract features from non-stationary ABA signals as an energy term, which is well-matched to defect status. In the front bogie, four vertical ABA signals of five seconds are measured from virtual accelerometers mounted on an axle box at the sampling rate of 2,000 Hz. The signals are further segmented at the time interval of single wheel rotation. Under the conditions of a wheel radius of 0.5 m and a train speed of 60 km/h, the number of wheel revolutions per second is approximately 5.3 Hz and varies slightly depending on the height of a wheel flat. Each ABA signal for one flat height is divided into 25 signal segments at the interval of a wheel revolution without overlap between signal segments. A total of 500 signal segments are prepared from a combination of four measurement positions and five flat heights as inputs for feature extraction. The extraction process consists of three steps in the time-frequency domain, namely, HT, WPD, and RMS. Figure 7 shows the structure of the proposed energy-based methodology.

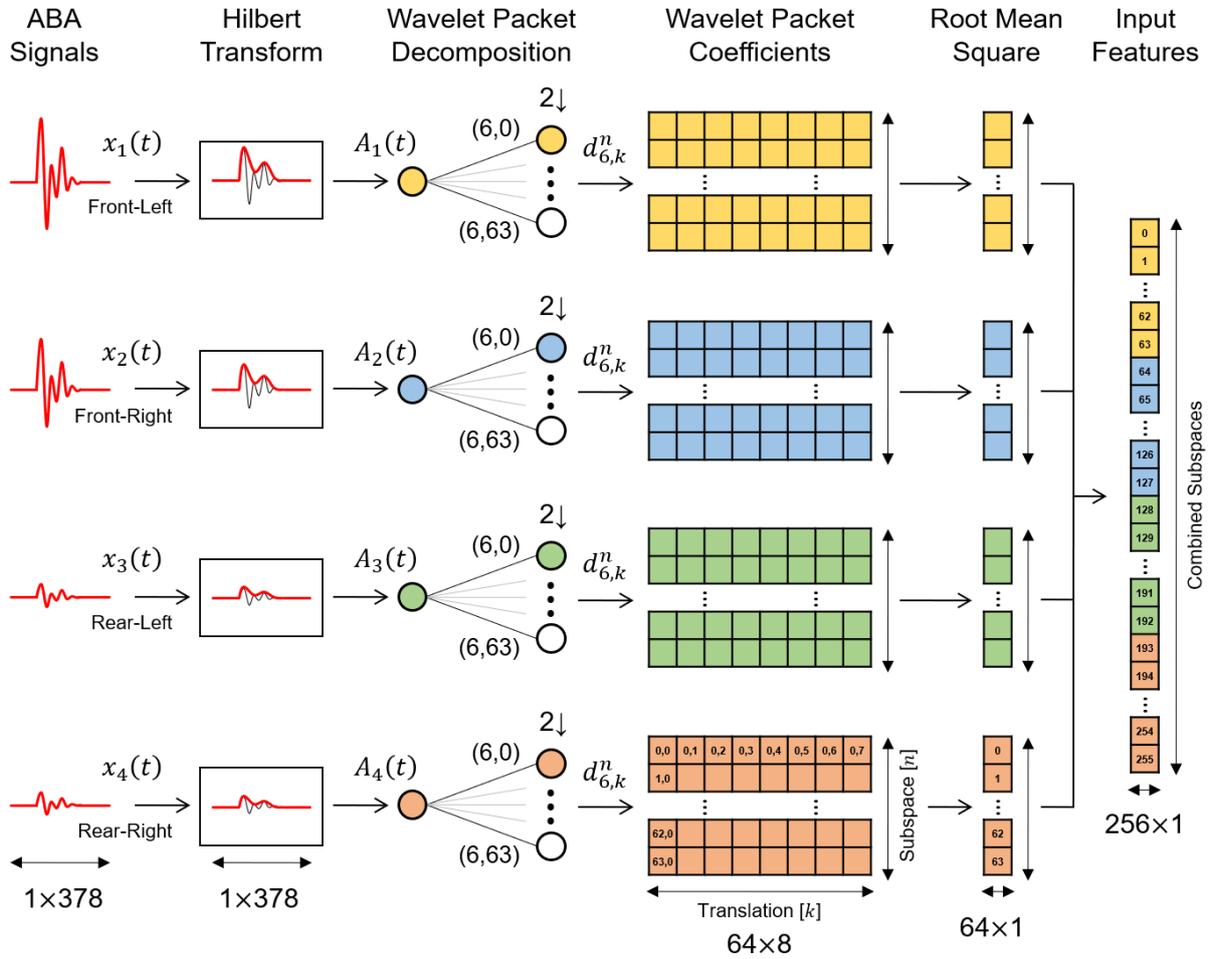

Figure 7 Structure of the proposed feature extraction method for extracting input features from four vertical ABA signals

In the first step, HT is applied to signal segments to extract the amplitude modulation part of ABA signals that is considered to be closely related to the height of a wheel flat in existing studies [26, 27]. Then, WPD is used to decompose the amplitude modulation part into orthogonal multiple subspaces on a balanced binary tree. The number of subspaces at the end of a binary tree is determined to be between 1 and 64 by decomposition level from 0 to 6. Each subspace contains wavelet packet coefficients as a decomposed and subsampled part of ABA signals. The data length is halved recursively during the decomposition process, but the sum of energy for all the subspace is always constant without loss. The RMS distributions of four ABA signals are calculated as an energy term from wavelet packet coefficients of subspaces at the end of a balanced binary tree and combined into one column as input features for defect prediction.

The above extraction process is applied repeatedly to all the signal segments. The extraction process of the amplitude modulation part from ABA signals is introduced in Section 3.2. The signal decomposition process is further described in Section 3.3.

## 3.2 Hilbert Transform

HT is a linear operator to transfer a real-valued signal $x_p(t)$ to a complex-valued signal $x_p^A(t)$ by convolution with the Dirac delta function of $1/\pi t$ [34], which is defined by:

$$H[x_p(t)] = \frac{1}{\pi} \int_{-\infty}^{\infty} \frac{x_p(\tau)}{t-\tau} d\tau \qquad \text{Eq. 3}$$

where $p$ is the measurement location in the front bogie, $t$ is time, $\tau$ is a dummy variable, and $H[x_p(t)]$ is the HT of a real-valued signal $x_p(t)$. As an analytic representation, a complex-valued signal $x_p^A(t)$ is defined as:

$$x_p^A(t) = x_p(t) + jH[x_p(t)] = A_p(t)e^{j\phi(t)} \qquad \text{Eq. 4}$$

where $A_p(t)$ and $\phi(t)$ are the analytic amplitude and phase, respectively, and are given below:

$$A_p(t) = \sqrt{x_p^2(t) + H^2[x_p(t)]} \qquad \text{Eq. 5}$$

Since HT is a phase shifter of 90 degrees, the analytic phase $\phi(t)$ between real-valued $x_p(t)$ and complex-valued signals $x_p^A(t)$ is always 90 degrees independent of frequency. A real-valued signal $x_p(t)$ consists of frequency and amplitude modulation parts. In the analytic amplitude $A_p(t)$, the frequency modulation part disappears because sine and cosine functions cancel each other by the phase shift of 90 degrees, and then, only the amplitude modulation part remains. In this work, HT is applied to extract the amplitude modulation part from four vertical ABA signals as shown in Figure 7. The data length remains the same as the signal segment with 378 samples. The output of HT is used as the input of WPD.

## 3.3 Wavelet Packet Decomposition

In existing studies, WD and EMD are applied for the orthogonal extraction of the desired signal components and the improvement of SNR before extracting the features from the ABA signals [24-26]. WD has an unbalanced binary tree biased towards low frequency and has a low resolution in the high frequency region. In EMD, the decomposition order and boundaries can be changed as a new component is added to a signal. As an alternative, WPD is used to decompose a signal into orthogonal subspaces on a balanced binary tree structure as an extended form of WD [35]. The decomposition boundaries on the frequency domain are evenly divided according to the level of WPD and not affected by changes in the components of a signal. The principle of WPD is defined in the form of a time-frequency function as:

$$W_{j,k}^n(t) = 2^{-j/2} W^n(2^{-j}t - k) \qquad \text{Eq. 6}$$

where $j$ is a scale parameter and the level of a binary tree structure, $n$ is a modulation parameter and the position from the left side of subspaces, and $k$ is a translation parameter and the position within the decomposed signal of a specific subspace $(j, n)$ on the binary tree structure. The scaling and wavelet functions of the Daubechies wavelets (db$A$ or d$N$) of length $N$ were used, which is one of the orthogonal wavelets and characterized by a maximal number $A$ of vanishing moments [36]. The Daubechies wavelet (db2 or d4) of length 4 with 2 vanishing moments was chosen as shown in Figure 8. The orthogonality between both functions means that they can be used as a quadratic mirror filter that has a magnitude response orthogonally mirrored from another filter.

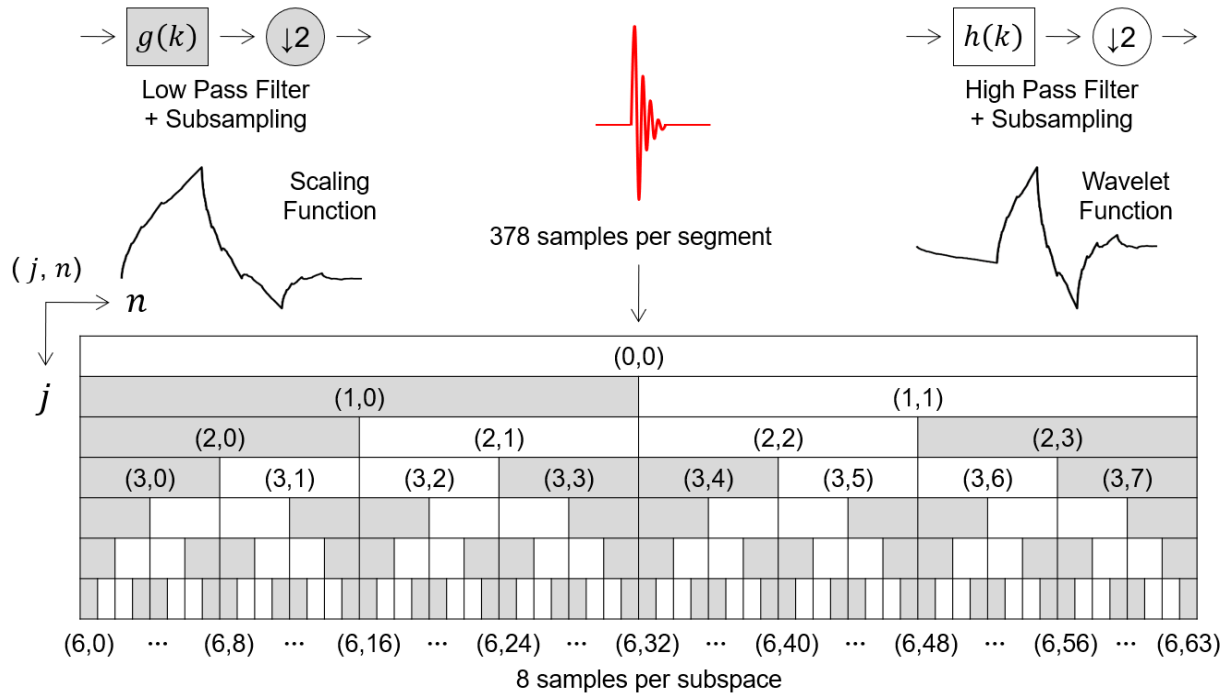

Figure 8 An example of the binary tree structure of 6-level WPD using Daubechies scaling and wavelet functions (db4 or d4)

Based on the quadratic mirror property, these functions operate as low- and high-pass filters of length $2N$, and then, a signal is decomposed into two orthogonal subspaces of equal bandwidth. The first two wavelet packet functions are the scaling and wavelet functions at the root level ($j=0$, $n=0$), which are defined by:

$$W_{0,0}^0(t) = W^0(t) \qquad \text{Eq. 7}$$

$$W_{0,0}^1(t) = W^1(t) \qquad \text{Eq. 8}$$

For each scale parameter $j$, the possible values of the modulation parameter $n$ are 0, 1, ..., $2^j - 1$. The remaining wavelet packet functions are defined by the following recursive relationships:

$$W_{j+1,k}^{2n}(t) = \sqrt{2} \int_{k=0}^{2N-1} g(k) W_{j,k}^n(2t-k) dt \qquad \text{Eq. 9}$$

$$W_{j+1,k}^{2n+1}(t) = \sqrt{2} \int_{k=0}^{2N-1} h(k) W_{j,k}^n(2t-k) dt \qquad \text{Eq. 10}$$

where $g(t)$ and $h(t)$ are the filter coefficients of low- and high-pass filters, respectively, and the wavelet packet coefficient $d_{j,k}^n$ is obtained as the inner product between the analytic amplitude of ABA signals $A_p(t)$ and wavelet packet function $W_{j,k}^n(t)$, as below:

$$d_{j,k}^n = \langle A_p(t), W_{j,k}^n(t) \rangle = \int A_p W_{j,k}^n(t) dt \qquad \text{Eq. 11}$$

Figure 8 shows the balanced binary tree structure of the 6-level WPD as an example. After applying HT, the analytic amplitude of ABA signals $A_p(t)$ is orthogonally decomposed into two subspaces, which are defined as the approximate (grey) and detail (white) coefficients in the low- and high-frequency subspaces, respectively. The quadratic filters $g(t)$ and $h(t)$ are applied recursively to filtered signals. The frequency range and the order of each subspace are determined symmetrically on the balanced binary tree structure. During the filtering process, since half the information is lost in the frequency domain, filtered signals are subsampled with the operator $\downarrow 2$. At the 6-level WPD, the number of subspaces is 64 from (6,0) to (6,63), and the length of the wavelet packet coefficient $d_{j,k}^n$ for each subspace is reduced from 378 samples to 8 samples by six iterations of subsampling. The matrix size of wavelet packet coefficients for each ABA signal is 64 by 8. Figure 9 shows the wavelet packet coefficients of decomposed ABA signals in the front bogie due to a flat height of 1e-1 mm, where a flat is located at the front-wheelset and left wheel. The magnitude and time of vibration responses are different depending on the position of measurement. In this work, WPD is applied to the amplitude modulation part of ABA signals from 0 to 6 levels. Then, wavelet packet coefficients at each subspace are used to calculate the RMS distribution as an energy term. The matrix size is reduced from 64 by 8 to 64 by 1. In Figure 10, the RMS distributions are compared by the position of ABA signals. One segment of input features is a combination of RMS distributions for four wheels. The shape of input features is 256 by 1.

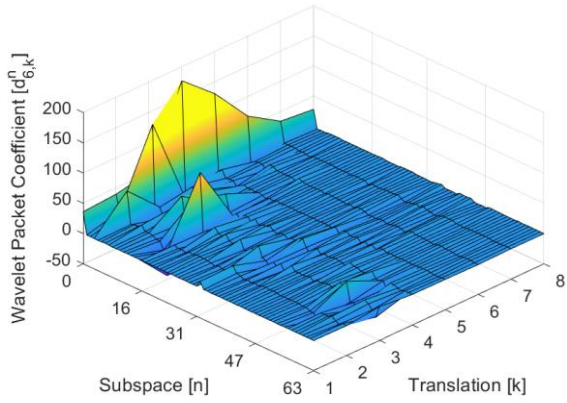 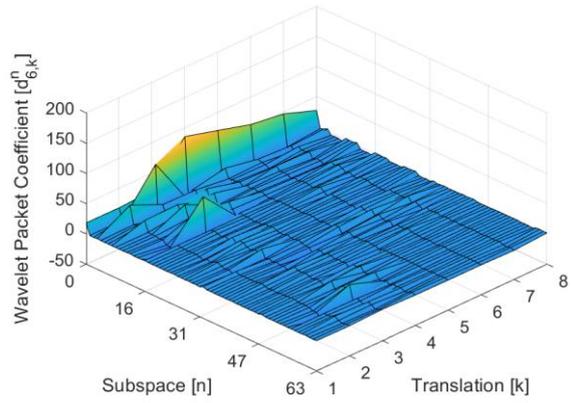

(a) Front Wheelset - Left Wheel (Defect)     (b) Front Wheelset - Right Wheel

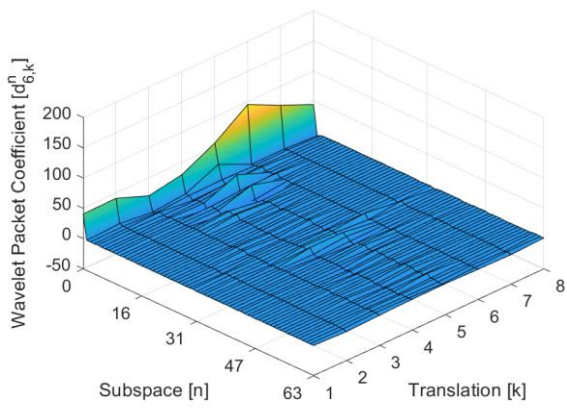 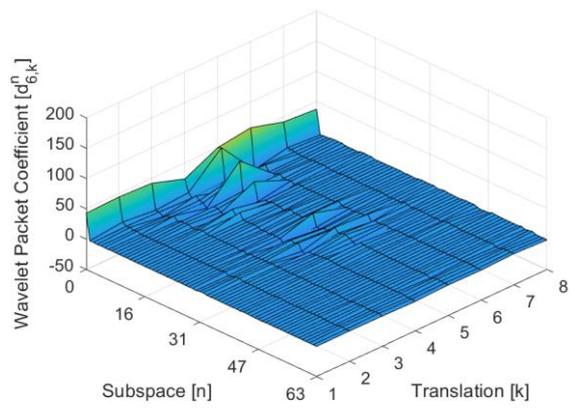

(c) Rear Wheelset - Left Wheel     (d) Rear Wheelset - Right Wheel

Figure 9 Comparison of the wavelet packet coefficients of decomposed ABA signals in the front bogie due to a flat height of 1e-1 mm at the 6-level WPD

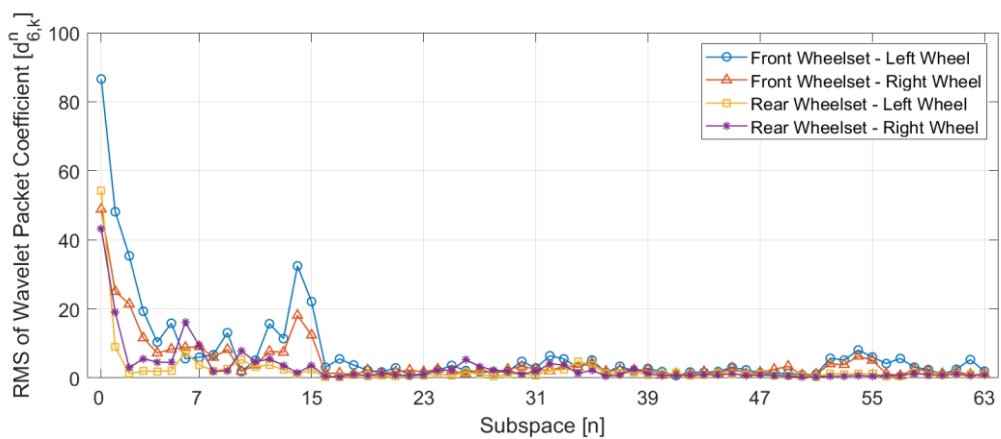

Figure 10 RMS distribution of wavelet packet coefficients by the measurement position of ABA signals at the 6-level WPD

# 4 Defect Prediction

## 4.1 Data Augmentation

A sufficient number of input features are required for the accuracy and robustness of a prediction model, but it is difficult to obtain all the necessary ABA signals through numerical simulations due to the computational load and time. As the height of a wheel flat approaches a critical value of 1e-0 mm, the magnitudes of ABA signals increase rapidly so that the widening gap between features can increase numerical errors in the predicted results of a defect prediction model. For this reason, data augmentation is performed to supplement insufficient input features from linear interpolation between original input features, which is one of the frequently used methods in preparing image datasets for machine learning [37].

Figure 11 shows the concept of data augmentation from original input features. For five flat heights (1e-4, 1e-3, 1e-2, 1e-1, and 1e-0 mm), 25 signal segments per flat height are obtained through the segmentation of ABA signals in five seconds. A total of 125 signal segments becomes original input features with a matrix size of 64 by 125 after the feature extraction process at the 6-level WPD and then is expanded to a matrix size of 256 by 125 for four measurement positions within a bogie. The number of rows varies from 4 by 125 to 256 by 125 depending on the WPD level from 0 to 6. The augmented input features are obtained from linear interpolation between two different sets of 25 signal segments at four interpolation points and both end points within two adjacent flat heights. At each interpolation or end point, the number of augmented input features is 625 by the Cartesian product of two different sets.

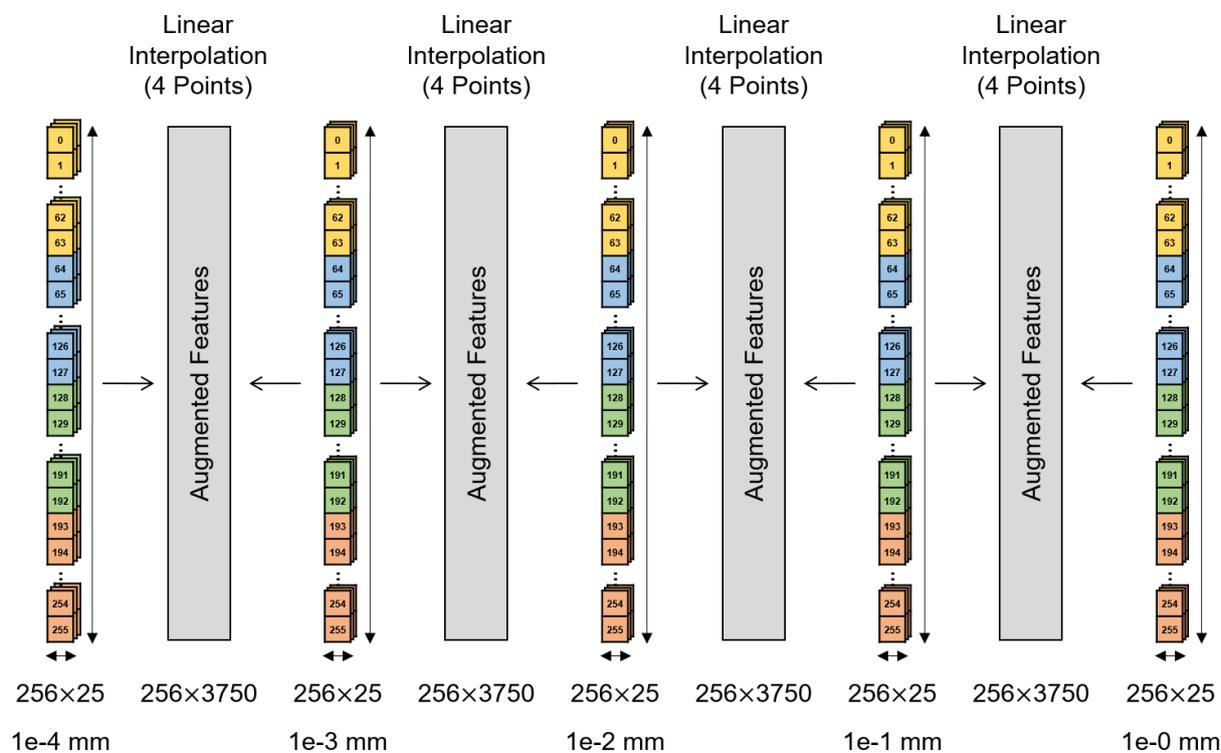

Figure 11 Data augmentation of original input features through linear interpolation

|   | s0 | s1 | s2 | s3 | ... | s252 | s253 | s254 | s255 |   |   | FL | FR | RL | RR |
|---|---|---|---|---|---|---|---|---|---|---|---|---|---|---|---|
| 0 | 7.776 | 1.346 | 0.409 | 1.647 | ... | -0.496 | -0.453 | -0.451 | 0.423 |   | 0 | 1e-0 | 0.00 | 0.00 | 0.00 |
| 1 | 4.112 | 3.254 | 1.818 | 1.073 | ... | -0.425 | -0.332 | -0.394 | -0.470 |   | 1 | 0.00 | 1e-0 | 0.00 | 0.00 |
| 2 | 7.384 | 0.556 | 0.923 | 3.084 | ... | -0.513 | -0.538 | -0.548 | -0.496 |   | 2 | 0.00 | 0.00 | 1e-0 | 0.00 |
| 3 | 3.965 | 1.775 | 0.879 | 1.396 | ... | -0.314 | -0.047 | -0.200 | -0.243 |   | 3 | 0.00 | 0.00 | 0.00 | 1e-0 |
| 4 | 7.354 | 3.587 | 3.833 | 2.794 | ... | -0.253 | -0.003 | -0.330 | -0.093 |   | 4 | 1e-0.6 | 0.00 | 0.00 | 0.00 |
| 5 | 6.432 | 3.327 | 2.225 | 1.688 | ... | -0.352 | -0.3717 | -0.385 | -0.370 |   | 5 | 0.00 | 1e-1.4 | 0.00 | 0.00 |
| 6 | 4.272 | 0.495 | 0.063 | 0.010 | ... | -0.297 | -0.354 | -0.195 | -0.191 |   | 6 | 0.00 | 0.00 | 1e-2.6 | 0.00 |
| 7 | 7.068 | 0.907 | 0.383 | 0.873 | ... | -0.305 | -0.289 | -0.362 | -0.359 |   | 7 | 0.00 | 0.00 | 0.00 | 1e-3.4 |

(a) Input features  (b) Output labels

Figure 12 An example of augmented input features and output labels

For a specific flat height, a total of 3,750 signal segments is obtained at six points within two adjacent flat heights. The same process is repeatedly applied to four gaps between five flat heights and four defect locations. After data augmentation, the number of signal segments increases from 125 to 60,000. As a result, the matrix size of augmented input features varies from 4 by 60,000 to 256 by 60,000 depending on the WPD level from 0 to 6. The augmented output labels are also obtained through the same process as above with the shape of 4 by 60,000. Figure 12 shows an example of augmented input features and output labels.

### 4.2 Training of Feedforward Neural Network

A neural network-based defect prediction model was created to define the non-linear relationship between input features and output labels. The FNN mimics interconnecting neurons mathematically in the form of a biologically inspired adaptive system. The structure of FNN consists of three types of layers, namely, the input, hidden, and output layers [38]. The number of neurons in the input layer equals the number of input features and varies from 4 to 256 depending on the WPD level from 0 to 6. For faster learning and convergence, input features are zero-centred by mean subtraction and normalized with a standard deviation of 1. The hidden layer is the core part of a defect prediction model in the form of a black box with hyperbolic tangent functions. The structural complexity is empirically determined after statistically comparing learning curves and residual errors of trained neural networks. In the first and second hidden layers, the number of neurons is chosen as 32 and 16, respectively. The output layer has four neurons corresponding to flat heights from 1e-4 mm to 1e-0 by the position of ABA signals in the front bogie on the output labels. The training process is performed to converge weight and bias values with the scaled conjugate gradient algorithm, which is one of the supervised learning algorithms, in the direction of minimizing the mean squared error of a cost function that is defined as the difference between actual and predicted labels. The neural network training tool provided by MATLAB is used.

Table 1 Comparison of the prediction accuracy of a trained feedforward neural network for detection and localization at different flat heights and WPD levels

(a) Detection

| Flat Height (Centre Value) [mm] | WPD j-level | | | | | | |
|---|---|---|---|---|---|---|---|
| | L0 | L1 | L2 | L3 | L4 | L5 | L6 |
| 1e-0 | 0.623 | 0.865 | 0.938 | 0.952 | 0.978 | 0.983 | 0.985 |
| 1e-1 | 0.813 | 0.871 | 0.918 | 0.942 | 0.983 | 0.991 | 0.994 |
| 1e-2 | 0.942 | 0.924 | 0.951 | 0.969 | 0.985 | 0.989 | 0.994 |
| 1e-3 | 0.730 | 0.891 | 0.945 | 0.960 | 0.980 | 0.987 | 0.993 |
| 1e-4 | 0.359 | 0.883 | 0.884 | 0.910 | 0.949 | 0.965 | 0.974 |
| Average | 0.693 | 0.877 | 0.927 | 0.947 | 0.975 | 0.983 | 0.988 |

(b) Localization

| Defect Location (Wheelset-Wheel) [-] | WPD j-level | | | | | | |
|---|---|---|---|---|---|---|---|
| | L0 | L1 | L2 | L3 | L4 | L5 | L6 |
| Front-Left | 0.748 | 0.879 | 0.925 | 0.946 | 0.972 | 0.983 | 0.986 |
| Front-Right | 0.753 | 0.890 | 0.928 | 0.949 | 0.974 | 0.982 | 0.988 |
| Rear-Left | 0.745 | 0.895 | 0.927 | 0.956 | 0.973 | 0.981 | 0.987 |
| Rear-Right | 0.750 | 0.881 | 0.928 | 0.945 | 0.972 | 0.982 | 0.988 |
| Average | 0.749 | 0.886 | 0.927 | 0.949 | 0.973 | 0.982 | 0.987 |

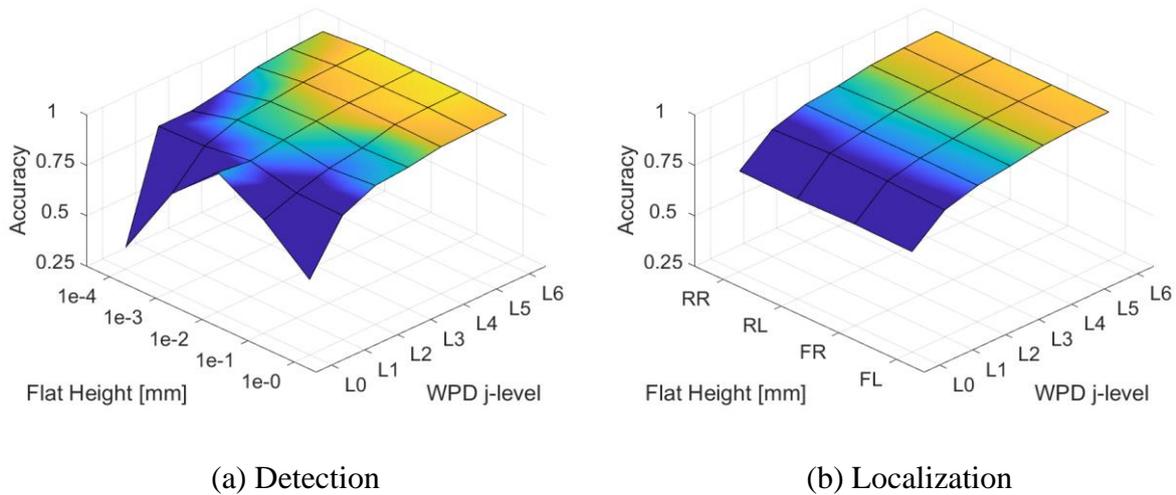

(a) Detection                 (b) Localization

Figure 13 Comparison of the surface of prediction accuracy for detection and localization at different flat heights and WPD levels

### 4.3 Performance Evaluation and Limitations

The performance of a trained defect prediction model was evaluated in terms of the detection and localization of a single wheel flat. The maximum error values of four predicted flat heights were used to calculate the prediction accuracy of detection and localization. For the defect prediction model trained with the original feature data, the accuracy of detection and localization is low regardless of the flat height and WPD level. After applying the data augmentation, residual errors in the cost function are reduced, and the prediction accuracy gradually increases to a value close to 1 as the WPD level increases. Table 1 and Figure 13 show the effects of the WPD level on the prediction accuracy of detection and localization. The zero-level WPD means no signal decomposition as the root level of a binary tree structure. The combination of the data augmentation and WPD 6-level show the best result for the detection and localization of a single wheel flat within one bogie. The results of using a higher level of WPD are similar to those of the WPD 6-level. In conclusion, the performance of a neural network-based prediction model can be improved by applying a higher frequency resolution of WPD and data augmentation to the RMS distribution of non-stationary ABA signals due to a single wheel flat.

Although the high prediction accuracy of detection and localization means that the RMS distribution as an energy term contains sufficient defect information, only a constant speed of 60 km/h and a single wheel flat are considered in the current train dynamics model. To validate the proposed feature extraction method and defect prediction model, the train dynamics model should consider further various train speeds, multiple wheel flats, and other vibration sources from rail and track models. These factors will negatively affect the performance of a defect prediction model because there are many other components as noisy as ABA signals in addition to the desired components closely related to wheel flats. Therefore, the future study will require the use of signal denoising techniques to eliminate unnecessary noise components for high SNR.

# 5  Conclusions

This study aims to identify the height and location of a single wheel flat from non-stationary ABA signals within one bogie. The train dynamics model with flexible wheelsets is used to generate ABA signals under the general specifications of MEMS capacitive accelerometers mounted on axle box. The modelling, simulation, and post-processing are automated using SIMULIA Simpack Rail and Python. The RMS distribution of ABA signals on a balanced binary tree is considered as orthogonal energy features for defect prediction and obtained by applying HT and WPD sequentially. The defect prediction model is created using FFN. Insufficient input features for the training process of a neural network are supplemented by the data augmentation. The performance of a trained defect prediction model is evaluated in terms of the prediction accuracy for the detection and localization of a single wheel flat. The results are as follows:

1. The proposed feature extraction method considers only four ABA signals within one bogie as a signal source for defect prediction. The vibrational energy caused by a single wheel flat is not easily transferred to other bogies due to high damping on structural transfer paths.

2. The bandwidth and acceleration range of interest are chosen to be up to 1,000 Hz and $\pm 200$ g and are sufficient for onboard structural health monitoring of wheel flats. The energy distribution of ABA signals is located mainly under 500 Hz, and the maximum amplitude of ABA signals is approximately $\pm 100$ g.

3. The performance of a trained defect prediction model is improved by the data augmentation and high level of WPD. The best results are obtained at the WPD 6-level. The prediction accuracy of detection and localization is close to 1 with low residual errors of a cost function.

In a future study, various train speeds, multiple wheel flats, and other vibration sources from rails and tracks should be included to generate more realistic vibration responses from a train dynamics model and to further validate the proposed feature extraction method.

## Acknowledgements

During this work, André Schmidt provided advice for modelling and simulation of train dynamics models. This work is supported by Universität Stuttgart, Chinesisch-Deutsches Forschungs- und Entwicklungszentrum für Bahn- und Verkehrstechnik Stuttgart, and Hefei University.

## Funding

The authors disclosed receipt of the following financial support for the research, authorship, and/or publication of this article:

## Declaration of conflicting interests

The authors declare that there are no conflicts of interest.

# References

[1] Chong S. Y., Lee J. R., Shin H. J. (2010). A review of health and operation monitoring technologies for trains. Smart Structures and Systems, 6(9), 1079-1105.

[2] Pieringer A., Kropp W., Nielsen J. C. O. (2014). The influence of contact modelling on simulated wheel/rail interaction due to wheel flats. Wear, 314(1-2), 273-281.

[3] CEN (European Committee for Standardization). (2016). Railway applications - In-service wheelset operation requirements - In-service and off-vehicle wheelset maintenance. EN 15313: 2016.

[4] Spiroiu M. A., Nicolescu M. (2018). Failure modes analysis of railway wheel. Proceedings of the 22th International Conference on Innovative Manufacturing Engineering and Energy, IManE&E 2018.

[5] Liu X. Z., Ni Y. Q. (2018). Wheel tread defect detection for high-speed trains using FBG-based online monitoring techniques. Smart Structures and Systems, 21(5), 687-694.

[6] Thompson D. (2008). Railway Noise and Vibration: Mechanisms, Modelling and Means of Control. Elsevier Science. ISBN 9780080914435.

[7] Vyas N.S., Gupta A.K. (2006). Modeling Rail Wheel-Flat Dynamics. Engineering Asset Management. Proceedings of the 1st World Congress on Engineering Asset Management, WCEAM 2006.

[8] Uzzal R. U. A., Ahmed A. K. W., Rakheja S. (2009). Analysis of pitch plane railway vehicle-track interactions due to single and multiple wheel flats. Proceedings of the Institution of Mechanical Engineers, Part F: Journal of Rail and Rapid Transit, 223(4), 375-390.

[9] Alemi A., Corman F., Lodewijks G. (2016). Condition monitoring approaches for the detection of railway wheel defects. Proceedings of the Institution of Mechanical Engineers Part F: Journal of Rail and Rapid Transit, 231(8), 961-981.

[10] Fraga-Lamas P., Fernández-Caramés T. M., Castedo L. (2017). Towards the Internet of Smart Trains: A Review on Industrial IoT-Connected Railways. Sensors, 17(6), 1457.

[11] Milne D., Pen L. L., Watson G., Thompson D., Powrie W., Hayward M., Morley S., (2016). Proving MEMS Technologies for Smarter Railway Infrastructure. Procedia Engineering, 143, 1077-1084.

[12] Kalker J. J. (1982). A fast algorithm for the simplified theory of rolling contact. Vehicle System Dynamics, 11(1), 1-13.

[13] Fisette P., Samin J. C. (1994). A new wheel/rail contact model for independent wheels. Archive of Applied Mechanics, 64, 180-191.

[14] Netter H., Schupp G., Rulka W., Schroeder K. (1998). New aspects of contact modelling and validation within multibody system simulation of railway vehicles. Vehicle System Dynamics 22(sup1), 246-269.

[15] Chaar N. (2007). Wheelset Structural Flexibility and Track Flexibility in Vehicle-Track Dynamic Interaction. Doctoral dissertation, KTH Royal Institute of Technology, Sweden.

[16] Jin X. (2016). Experimental and numerical modal analyses of high-speed train wheelsets. Proceedings of the Institution of Mechanical Engineers, Part F: Journal of Rail and Rapid Transit, 230(3), 643-661.

[17] Wu X., Rakheja S., Ahmed A. K. W., Chi M. (2018). Influence of a flexible wheelset on the dynamic responses of a high-speed railway car due to a wheel flat. Proceedings of the Institution of Mechanical Engineers, Part F: Journal of Rail and Rapid Transit, 232(4), 1033-1048.

[18] Alexandrou G., Kouroussis G., Verliden O. (2016). A compresensive prediction model for vehicle/track/soil dynamic response due to wheel flats. Proc IMechE Part F: J Rail and Rapid Transit, 230(4), 1088-1104.

[19] Zhou C., Gao L., Xiao H., Hou B. (2002). Railway Wheel Flat Recognition and Precise Positioninng Method Based on Multisensor Arrays. Applied Sciences, 10(4), 1297.

[20] Jardine A. K. S., Lin D., Banjevic D. (2006). A review on machinery diagnostics and prognostics implementing condition-based maintenance. Mechanical Systems and Signal Processing, 20(7), 1483-1510.

[21] Li C., Luo S., Cole C., Spiryagin M. (2017). An overview: modern techniques for railway vehicle on-board health monitoring systems. Vehicle System Dynamics, 55(7), 1045-1070.

[22] Liang B., Iwnicki S. D., Zhao Y., Crosbee D. (2013). Railway wheel-flat and rail surface defect modelling and analysis by time–frequency techniques. Vehicle System Dynamics, 51(9), 1403-1421.

[23] Song Y., Liang L., Du Y., Sun B. (2020). Railway Polygonized Wheel Detection Based on Numerical Time-Frequency Analysis of Axle-Box Acceleration. Applied Sciences, 10(5), 1613.

[24] Jia S., Dhanasekar M. (2007). Detection of Rail Wheel Flats using Wavelet Approaches. Structural Health Monitoring, 6(2), 121-131.

[25] Krummenacher G., Ong C. S. Koller S., Kobayashi S., Buhmann J. M., (2018). Wheel Defect Detection With Machine Learning. IEEE Transactions on Intelligent Transportation Systems, 19(4), 1176-1187.

[26] Jiang H., Lin J. (2018). Fault Diagnosis of Wheel Flat Using Empirical Mode Decomposition-Hilbert Envelope Spectrum. Mathematical Problems in Engineering.

[27] Nowakowski T., Komorski P., Szymański G. M., Tomaszewski F. (2017). Wheel-flat detection on trams using envelope analysis with Hilbert transform. Latin American Journal of Solids and Structures, 16(1).

[28] Cao K., Sun F., Xing X. (2017). Safety region estimation and fault diagnosis of wheels based on LSSVM and PNN. AIP Conference Proceedings, 1820(1).

[29] Brizuel J., Fritsch C., Ibanez A. (2011). Railway wheel at detection and measurement by ultrasound. Transportation Research Part C: Emerging Technologies, 19(6), 975-984.

[30] Gabor D. (1946). Theory of communication. Journal of the Institute of Electrical Engineering, 93(3), 429-457.

[31] Craig R., Bampton M. (1968). Coupling of Substructures for Dynamic Analyses. AIAA Journal, 6(7), 1313-1319.

[32] Arnold M., Schiehlen W. (2009). Simulation Techniques for Applied Dynamics. Springer Verlag. ISBN 9783211895474.

[33] Analog Devices ADXL 375 (2014). MEMS capacitive accelerometer. Data sheet of digital MEMS accelerometer ADXL 375. Website: www.analog.com. Last visited: 15.06.2020.

[34] King F. W. (2009). Hilbert Transforms. Cambridge University Press. ISBN 9780511721458.

[35] Mallat S. G. (1989) A theory for multiresolution signal decomposition: the wavelet representation. IEEE Transactions on Pattern Analysis and Machine Intelligence, 11(7), 674-693.

[36] Daubechies I. (1990). The wavelet transform, time-frequency localization and signal analysis. IEEE Transactions on Information Theory 36(5), 961-1005.

[37] Shorten C., Khoshgftaar T. M. (2019). A survey on Image Data Augmentation for Deep Learning. Journal of Big Data, 6(60),

[38] Svozil D., Kvasnicka V., Pospichal J. (1997). Introduction to multi-layer feed-forward neural networks. Chemometrics and Intelligent Laboratory Systems, 39(1), 43-62.